# Modèles de jeux sérieux collaboratifs et mobiles

Iza Marfisi-Schottman, Sébastien George
LUNAM Université, Université du Maine, EA 4023, LIUM, 72085 Le Mans, France
iza.marfisi@univ-lemans.fr, sebastien.george@univ-lemans.fr

**Résumé.** Les jeux sérieux collaboratifs sur dispositifs mobiles intègrent tous les ingrédients nécessaires à motiver la nouvelle génération d'apprenants. Cependant, la conception d'un scénario de jeu cohérent, qui combine ressorts ludiques, mobilité, collaboration et activités pédagogiques, est très complexe. Dans cet article, nous proposons trois modèles de jeux, qui intègrent naturellement ces caractéristiques, de façon à favoriser l'atteinte des objectifs pédagogiques.

**Mots-clés.** Jeux sérieux, apprentissage collaboratif, apprentissage situé

**Abstract.** Mobile Collaborative Learning Games combine all the ingredients necessary to attract students' attention and engage them in learning activities. However, designing a coherent game scenario that combines mobility, game mechanics and collaborative learning is quite a challenge. In this article, we propose three game patterns that naturally integrate these mechanics in order to support educational goals.

## 1    Collaborer, jouer et bouger pour mieux apprendre

Les enseignants sont contraints de faire évoluer leurs pratiques afin de motiver la nouvelle génération d'élèves [1]. Dans cette première partie de l'article, nous allons décrire trois types d'activités pédagogiques qui sembles tous particulièrement prometteuses et qui se combinent avantageusement.

Tout d'abord, les **activités collaboratives** sont souvent utilisées par les enseignants [2]. Ceci parait tout à fait naturel étant donné que le travail en groupe facilite l'apprentissage par les *interactions sociales* et augmente l'*engagement personnel* des apprenants [3]

Récemment, le concept de jeux éducatifs s'est également développé pour motiver la nouvelle génération d'élèves. L'idée principale est d'utiliser des **activités ludiques,** et plus précisément les ressorts du jeu, tel que la compétition ou les récompenses de façon à captiver l'attention des apprenants et leur faire adopter une *place active* dans leur processus d'apprentissage [4]. Quand elles sont utilisées correctement, les activités ludiques déclenchent également des *émotions*, qui ont un effet positif sur l'engagement et la mémoire [5]. De plus, la dernière génération de jeux éducatifs jouit des aspects *attractifs* et *immersifs* des technologies du jeu vidéo.

Enfin, l'utilisation d'**activités mobiles** pour l'enseignement s'est également accrue maintenant que la majorité des jeunes possède un smartphone ou une tablette (75% des 18-24 ans en 2013 [6]). La mobilité permet tout d'abord l'*apprentissage situé*, pendant des visites de sites archéologiques [7] par exemple ou pendant que l'étudiant répare de vrais moteurs de voiture [8]. De plus, la mobilité peut aussi être très utile pour expliquer des concepts avec une *simulation participative*. Par exemple, dans le jeu *Disease Simulation* [9], qui enseigne le phénomène de propagation de virus, les enfants portent des voyants lumineux qui indiquent s'ils sont infectés et propagent le virus quand ils s'approchent d'autres enfants.

### 1.1 Synergies entre types d'activités pédagogiques

Bien que les activités **collaboratives**, **ludiques** et **mobiles** puissent améliorer l'apprentissage séparément, nous pensons que leur combinaison pourrait être encore plus efficace grâce aux synergies qui les lient. Comme nous l'avons vu, certains chercheurs ont déjà combiné ces types d'activités avec succès [7][9], mais il n'existe pas encore d'études qui identifient clairement en quoi ce trio est particulièrement efficace. Nous proposons donc, dans le prochain paragraphe, d'analyser les synergies entre la collaboration, les mécaniques de jeu et la mobilité, représentées par les flèches en pointillés de la
Fig. **1**. Les flèches pleines représentent les effets de ces mécaniques sur l'apprentissage, décrites précédemment.

**Collaboration et Jeu**
La mise en place d'activités collaboratives pour l'apprentissage, dans lesquels tous les participants contribuent, est très complexe. Afin d'inciter les membres d'une équipe à accomplir leurs tâches, Gomez *et al.* [10] ont développé un jeu pour trois participants, dans lequel chaque coéquipier manipule une souris, connectée au même écran. Pour gagner le jeu, l'équipe doit accomplir un ensemble de tâches qui ne peuvent être effectuées en utilisant qu'une seule des souris. En d'autres termes, les équipiers disposent de différents outils (dans ce cas, une souris), qui leur permettent d'accomplir des tâches complémentaires et ainsi atteindre des objectifs de façon collaborative. Nous sommes convaincus que les mécaniques de jeu offrent un moyen naturel de mettre en place de telles situations. En effet, comme dans le jeu *Environmental Detective* [11]*,* le scénario peut être conçu de telle façon que les joueurs se voient affectés *des rôles avec des objectifs complémentaires* (ex. employé d'une entreprise de construction, militant écologiste), forçant ainsi les apprenants à défendre leurs points de vue pour gagner le jeu. Le jeu peut également pousser ce principe encore plus loin en donnant des pouvoirs ou des *outils spécifiques* à chaque rôle (ex. troll, sorcier, alchimiste) de façon à ce que certains apprenants soient les seuls à pouvoir accomplir des tâches.
   Inversement, l'utilisation de la collaboration est reconnue comme une mécanique ludique [12] grâce aux *interactions sociales* et aux *défis supplémentaires* engendrés par la prise de décision collaborative.

**Jeu et Mobilité**
La mobilité ouvre de nouvelles possibilités pour enrichir l'expérience des joueurs en utilisant des *objets réels* (ex. plantes, bâtiments, animaux) dans leur *contexte naturel* (ex. site archéologique ou géologique, entreprise, forêt). De plus, plusieurs études ont montré que *l'excitation physiologique* engendrée quand un joueur bouge pendant le jeu, augmente son taux d'engagement [13]. Enfin, les fonctionnalités telles que le GPS ou l'accéléromètre collectent des *feedbacks* utiles pour adapter le jeu. Par exemple, si le joueur est coincé dans la même zone, le jeu peut lui donner un indice.

Inversement, les récompenses dans le jeu, telles que des points supplémentaires ou un déblocage de l'histoire, peuvent ***motiver l'apprenant à accomplir une activité dans le monde réel***. Ceci est par exemple le cas dans *Rewild*[1], un jeu d'aventure qui motive les randonneurs débutants à gravir des montagnes.

**Mobilité et Collaboration**
Si les activités collaboratives sont conçues de façon à ce que les joueurs doivent synchroniser leurs déplacements pour couvrir toute la grille de jeu par exemple, elles apportent une dimension ***plus amusante*** à la mobilité.

Inversement, la mobilité peut aboutir à des situations dans lesquelles les membres de l'équipe sont physiquement dispersés et n'ont pas les moyens de communiquer. Ce type de situations rend les activités collaboratives beaucoup ***plus stimulantes***. De plus, lors d'un travail collaboratif, il peut arriver que les coéquipiers ne progressent pas à la même vitesse. Avec l'utilisation d'une ***interface mobile individuelle***, chaque apprenant peut naviguer librement dans ses ressources et soumettre ses contributions quand il se sent prêt.

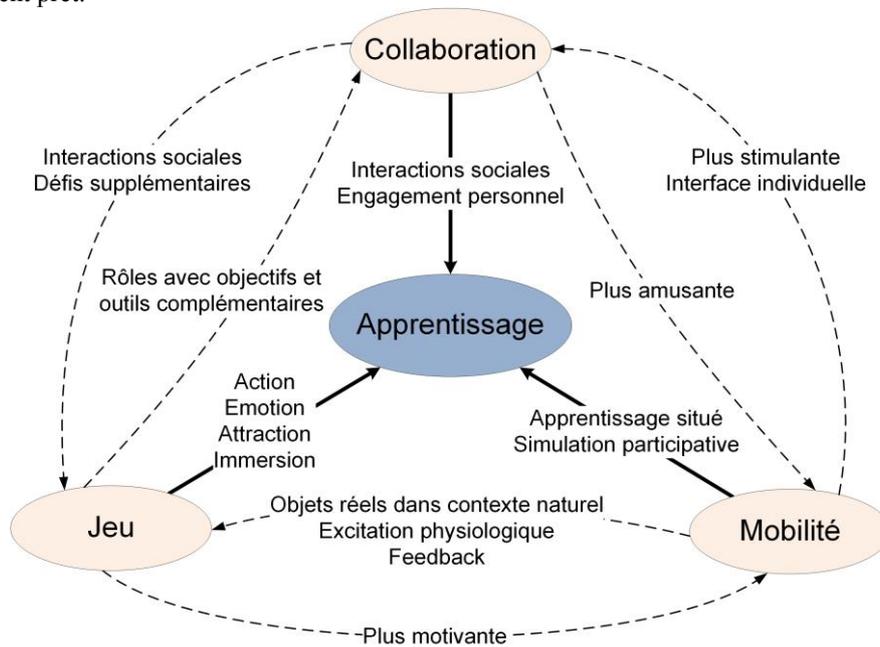

**Fig. 1.** Synergies entre la collaboration, le jeu et la mobilité

---

[1] http://www.rewild.fr

## 2 Conception de Jeux Sérieux Collaboratifs et Mobiles

Comme nous l'avons vu, plusieurs applications éducatives tirent déjà avantage des synergies entre les activités collaboratives, ludiques et mobiles. Cependant, leur conception demeure expérimentale, sans méthode pour garantir une synergie maximale. Afin de faciliter leur conception, nous proposons, dans la suite, trois patrons de jeu qui intègrent la collaboration, le jeu et la mobilité de façon à obtenir ces synergies et qui supportent une grande variété d'objectifs pédagogiques.

### 2.1 Patron « Jeu de Rôle Grandeur Nature »

Les jeux de rôle Grandeur Nature (GN) sont construits autour d'activités de jeu de rôles, guidés par des règles. Les joueurs choisissent les actions du personnage qu'ils incarnent en fonction des objectifs à atteindre et des alliances avec les autres personnages, créant ainsi une infinité d'histoires possible pour le jeu. Les GN sont habituellement orchestrés par un maître du jeu, qui facilite le déroulement du jeu et tient note des scores. D'un point de vue éducatif, les GN semblent parfaitement adaptés pour améliorer les aptitudes sociales (ex. communication, négociation) et comprendre des sujets complexes qui impliquent plusieurs acteurs. De plus, le rôle du maitre de jeu est parfait pour les enseignants parce qu'il leur permet d'observer les apprenants et d'adapter le jeu. Une variante consiste à réduire les rôles au minimum pour se concentrer sur les règles d'interaction entre joueurs. Ce type de jeu peut être utilisé pour créer des simulations participatives pour enseigner des mécanismes, tels que la propagation d'une maladie, comme dans le jeu *Disease Simulation* [9].

Dans la suite, nous décrivons chaque patron de JSCM avec un exemple et une structure spécifique, tel que les patrons de jeu proposés par [12]. Ces exemples sont inspirés de vrais jeux conçus et utilisés par des enseignants[2].

**Exemple 1.** Exemple du patron « jeux de rôle grandeur nature » – Construction de voie rapide

> **Objectif pédagogique.** Le jeu permet d'enseigner les mécanismes de la démocratie et d'entrainer les élèves à défendre leurs points de vue en collectant des données et en les présentant à l'oral.
> **Jeu**. Le maire a pour projet de construire une voie rapide à travers une portion de la ville, mais il veut d'abord en débattre avec les habitants.
> **Mobilité**. Avec un plan de la zone de construction, les joueurs doivent aller sur le terrain pour trouver des éléments qui sont en faveur ou contre la construction de la route. Régulièrement, les joueurs sont conviés à une assemblée générale pour défendre leurs points de vue sur le projet.
> **Collaboration**. Les joueurs ont des rôles complémentaires de façon à représenter la population de la ville :
> - le maire qui voudrait que sa ville prospère, mais ne veut pas aller à l'encontre de l'avis de la population
> - les citoyens qui réduiront leur temps de trajet quotidien grâce à cette nouvelle route
> - les commerçants et entrepreneurs du BTP qui accroitraient leur chiffre d'affaires
> - les militants qui veulent protéger le parc naturel à travers lequel la route doit passer
> - deux journalistes, un pour et l'autre contre la construction de la route, qui doivent mener des sondages et tenir la population informée des derniers rebondissements liés au projet
> Les mécaniques de score sont différentes pour chaque rôle et sont conçues de façon à ce que les joueurs doivent négocier pour arriver à une solution qui les satisfait tous.

---

[2] Caroline Juneau-Sion, Patrick Prévôt et Carmelo Ardito

### 2.2    Patron « Jeu d'investigation »

Le scénario de jeu d'investigation est construit autour d'une énigme centrale que les joueurs doivent résoudre en collectant et analysant des. Ce modèle est très similaire à la méthode d'apprentissage à base de cas. En effet, celle-ci consiste à poser un problème aux apprenants, inspirés d'une situation réelle, en les plaçant dans la position du preneur de décision. Nous pensons que ce modèle de jeu est donc parfaitement adapté pour entrainer les apprenants à utiliser leurs connaissances théoriques sur des cas concrets. Ces cas peuvent couvrir une très grande variété de compétences telle que trouver la pathologie dont souffre un patient, identifier la cause d'une panne, ou encore trouver la meilleure solution pour un divorce.

**Exemple 2.** Exemple du patron « jeu d'investigation » – Trois heures pour sauver Kogatana

> **Objectif pédagogique.** Le jeu permet d'entrainer les apprenants à utiliser des outils de prise de décision (ex. brainstorming, grille de critères) sur un vrai cas complexe.
> **Jeu**. Les étudiants sont embauchés par une entreprise (Kogatana) pour déterminer ce qui cause la récente perte d'argent. Ils ont un temps limité pour identifier le problème et proposer une solution.
> **Mobilité**. Les joueurs doivent se rendre dans différentes salles du bâtiment pour mener leur investigation. En scannant les QR code dans les salles, ils obtiennent des documents et des vidéos des employés qui sont interviewés. L'enseignant rassemble régulièrement les élèves pour des sessions de débriefing pendant lesquelles ils discutent des informations recueillies et des outils qu'ils doivent utiliser pour les aider.
> **Collaboration.** Pour finir le jeu à temps, les joueurs, qui sont par groupe de trois, doivent impérativement s'organiser et se partager les tâches (ex. partage des salles à visiter).

### 2.3    Patron « Chasse aux trésors »

Un jeu de chasse aux trésors est un jeu dans lequel les joueurs cherchent un ou plusieurs objets cachés, à l'aide d'une suite d'indices. Parce que ce type de jeu incite les joueurs à explorer et s'approprier l'environnement physique, nous pensons qu'il est tout particulièrement adapté pour enseigner les caractéristiques d'objets réels dans leurs contextes naturels.

**Exemple 3**. Exemple du patron « chasse aux trésors » – La soirée de Bacchus

> **Objectif pédagogique.** Le jeu permet d'enseigner les us et coutumes de la Rome antique et les évènements qui ont conduit à la fin tragique de Pompéi.
> **Jeu**. Les joueurs sont embauchés par Bacchus pour organiser une grande soirée en son honneur. On leur fournit une liste d'objets qu'ils doivent trouver (boissons, pitances, invités et animation) et une bourse de pièces d'or. Cependant, la quantité d'argent est loin d'être suffisante et les joueurs devront donc faire un choix : acheter ce qu'ils peuvent et avoir un score très bas sur la jauge de satisfaction de Bacchus ou voler ! S'ils décident de voler, le volcan commencera à montrer des signes d'éruption et les habitants virtuels de Pompéi leur diront que les dieux désapprouvent les agissements immoraux de la population. À la fin du jeu, quelle que soit la décision prise par les joueurs, le volcan rentre en éruption : si les joueurs ont volé, les dieux qui déclenchent l'éruption et s'ils n'ont pas volé, c'est la rage de Bacchus qui déclenche l'éruption !
> **Mobilité**. Les apprenants ont quelques heures pour physiquement explorer les ruines de Pompéi et trouver les objets sur la liste au vrai endroit où les magasins se trouvés. Pour trouver les invités et l'animation, ils ont le choix entre la maison des poètes et danseurs, la baraque des gladiateurs et des villas de nobles.
> **Collaboration.** Les joueurs doivent se distribuer les taches pour collecter les objets à temps. Ils devront également se contacter, quand ils sont à court d'argent, pour décider de la stratégie à adopter.

## 3　Conclusion

Dans cet article, nous analysons comment les activités *collaboratives*, *ludiques* et *mobiles* peuvent motiver la nouvelle génération d'élèves et améliorer leurs processus d'apprentissage. Nous montrons également que, quand ces trois types d'activités sont combinés dans un JSCM (Jeu Sérieux Collaboratif et Mobile), ils créent des synergies et sont donc encore plus efficaces. Cependant, il n'y a pour l'heure, aucune méthode pour faciliter la conception de tels jeux. Afin de répondre à cette problématique, nous proposons trois patrons de JSCM : les *jeux de rôle grandeur nature*, les *jeux de mystères* et les *chasses aux trésors*. Ces modèles créent naturellement des synergies entre les activités collaboratives, ludiques et mobiles et peuvent s'adapter à une grande diversité d'objectifs pédagogiques. Nous travaillons actuellement sur des outils auteurs qui intègrent des coquilles de scénario, correspondant à ces trois patrons, afin que les enseignants puissent, eux-mêmes concevoir des jeux adaptés à leurs besoins et les exécuter sur les plateformes à leurs dispositions.